\documentclass[letter, 10pt, conference]{ieeeconf}      
\usepackage{graphicx}
\usepackage{amsfonts,amssymb,graphicx,enumerate}
\usepackage[centertags]{amsmath}
\usepackage{hyperref}
\usepackage[table,xcdraw]{xcolor}
\usepackage{theorem}
\usepackage{multirow}
\usepackage[utf8]{inputenc}
\usepackage{caption}

\usepackage{tikz}
\usetikzlibrary{automata,arrows,calc,positioning,intersections}

\tikzset{
    old inner xsep/.estore in=\oldinnerxsep,
    old inner ysep/.estore in=\oldinnerysep,
    double circle/.style 2 args={
        circle,
        old inner xsep=\pgfkeysvalueof{/pgf/inner xsep},
        old inner ysep=\pgfkeysvalueof{/pgf/inner ysep},
        /pgf/inner xsep=\oldinnerxsep+#1,
        /pgf/inner ysep=\oldinnerysep+#1,
        alias=sourcenode,
        append after command={
        let     \p1 = (sourcenode.center),
                \p2 = (sourcenode.east),
                \n1 = {\x2-\x1-#1-0.5*\pgflinewidth}
        in
            node [inner sep=0pt, draw, circle, minimum width=2*\n1,at=(\p1),#2] {}
        }
    },
    double circle/.default={3pt}{black}
}

\usepackage{boxedminipage, calc}
\definecolor{red}{rgb}{.840,.352,.125}
\definecolor{xlightred}{rgb}{.961,.855,.793}
\definecolor{green}{rgb}{.379,.605,.105}
\definecolor{xlightgreen}{rgb}{.875,.914,.824}
\definecolor{brown}{rgb}{.871,.762,.523}
\definecolor{lightbrown}{rgb}{.922,.855,.711}
\definecolor{xlightbrown}{rgb}{.969,.945,.895}
\definecolor{blue}{rgb}{.105,.379,.605}
\definecolor{lightblue}{rgb}{.672,.793,.875}
\definecolor{xlightblue}{rgb}{.824,.875,.914}
\definecolor{topgray}{gray}{.7}

\newcommand{\bydef}{\stackrel{\Delta}{=}}
\newcommand{\byplus}{\stackrel{+}{=}}

\newcommand{\bfy}{\mathbf{y}}
\newcommand{\bfx}{\mathbf{x}}
\newcommand{\bfu}{\mathbf{u}}

\newcommand{\ie}{\textit{i}.\textit{e}.}

\graphicspath{../figures/}
\newcommand{\y}{\mathbf{y}}
\newcommand{\x}{\mathbf{x}}
\newcommand{\z}{\mathbf{z}}
\newcommand{\bu}{\mathbf{u}}
\newcommand{\w}{\boldsymbol \omega}

\newcommand{\m}{\widehat{\x}}

\newcommand{\F}{\mathcal{F}}

\renewcommand{\t}[1]{\mathrm{T}#1}
\newcommand{\N}{\mathcal{N}}
\newcommand{\R}{\mathcal{R}}

\renewcommand{\d}[1]{\;\mathrm{d}#1}

\usepackage{algpseudocode}
\IEEEoverridecommandlockouts                              

\usepackage{fancybox}
\usepackage{caption}
\usepackage{subcaption}

\title{\LARGE \bf Identification of Piecewise Affine State-Space Models via Expectation Maximization*}

\author{Rafael Rui$^{1}$, Tohid Ardeshiri$^{2}$ and Alexandre Bazanella$^{1}$%
\thanks{*This work is supported by Swedish research council (VR), project scalable Kalman filters, Conselho Nacional de Desenvolvimento Científico e Tecnológico (CNPq), Centro de Pesquisa e Inovação Suéco-Brasileiro (CISB) and Saab AB}
\thanks{$^{1}$Rafael Rui and Alexandre Bazanella are  with Department of Electrical Engineering, Universidade Federal do Rio Grande do Sul, Porto Alegre 90040-060, Brazil {\tt\small  rafael.rui, bazanella@ufrgs.br}}%
\thanks{$^{2}$Tohid Ardeshiri is with the Department of Electrical Engineering, Link\"{o}ping University, 58183 Link\"{o}ping, Sweden,
        {\tt\small tohid@isy.liu.se}}%
}

\begin{document}
\maketitle

\begin{abstract}
This paper deals with the identification of piecewise affine state-space models. These models are obtained by partitioning the state or input domain into a finite number of regions and by considering affine submodels in each region. The proposed framework uses the Expectation Maximization (EM) algorithm to identify the parameters of the model. In most of the current literature, a discrete random variable with a discrete transition density is introduced to describe the transition between each submodel, leading to a further approximation of the dynamical system by a jump Markov model.
On the contrary, we use the cumulative distribution function (CDF) to compute the probability of each submodel given the measurement at that time step. Then, given the submodel at each time step the latent state is estimated using the Kalman smoother. Subsequently, the parameters are estimated by maximizing a surrogate function for the likelihood. The performance of the proposed method is illustrated using the simulated model of the JAS 39 Gripen aircraft.
\end{abstract}


\section{Introduction}
\label{sec:intro}

Our goal in this paper is the identification of piecewise affine state-space (PWASS) models. The PWASS models are the next natural step in the approximation of nonlinear state-space models (SSM); Instead of linearizing  a nonlinear dynamical system by a single linear SSM, one can divide it into several affine submodels turning the identification of a nonlinear system into the identification of several affine submodels. 
 PWASS models are a particular case of piecewise affine (PWA) models.  Such models are used to approximate nonlinear dynamical systems and have been considered in several fields, such as automatic control \cite{ChzeEngSeah2009a}, signal processing \cite{Doucet2001} and computer vision \cite{Vidal2006}. They are obtained by partitioning the state or the input domain into a finite number of polyhedral regions; and by considering an affine submodel in each region. 

The identification problem of PWA models is a challenging problem that involves the estimation of both the parameters of the affine submodels and the coefficients that define the boundaries of each region of the state domain. The tutorial paper  \cite{Paoletti2007} discusses the main issues and difficulties connected to hybrid systems identification.

The majority of the identification methods proposed in the literature focus on the particular case of piecewise affine autoregressive with exogenous inputs (PWARX) models \cite{Ohlsson2013,Hartmann2015,MihalyPetreczky2012}. A PWASS model can be written as a PWARX model, and the equivalence between them is well-known \cite{Weiland2006}. The authors in \cite{Weiland2006} have shown that any observable PWASS model admits a representation as a PWARX model. It has been observed that PWARX systems are strictly contained in the class of PWASS models and that the number of submodels (and thus the number of parameters) might grow considerably when a PWASS model is converted into a minimum-order equivalent PWARX representation. On the other hand, given a PWARX model with $N_{r}$ submodels, it is possible to find an equivalent PWASS model with $N_{r}$ submodels, but the equivalent PWASS model will not necessarily  be a minimal realization \cite{Weiland2006}. 

Another reason that makes the PWASS representation be more appealing compared to PWARX representation is that the majority of existing hybrid and piecewise analysis and control methods such as \cite{Sontag1981,Petreczky2010} are based on SSM; and also SSM are more suitable to deal with multiple input and multiple output (MIMO) systems \cite{Bako2013}.

The identification of PWASS models undergoes some identifiability problems \cite{Vidal2002, Petreczky2010}. If one assumes that either the state is not fully measurable or that one does not fix some parameters, then the model will suffer from a realization problem, i.e., one can only determine the model up to a linear state transformation. For that reason, the few works that deal with the identification of PWASS models have to make some assumption such as: minimum dwell-time assumption  \cite{Borges2005}, observability and controllability  assumptions for each submodel \cite{Laurent2009}, known switching times assumption \cite{Verdult2004}, and full state knowledge assumption \cite{Bako2013}. 

 In most of the current literature, a discrete random variable with a discrete transition density is introduced to describe the transition among the different regions, leading to a  further approximation of the dynamical system by a jump Markov model \cite{Fox2011}. That is, the PWASS model is seen as an extension of hidden Markov models (HMMs) in which each HMM state is associated with a linear dynamical process. Several nonlinear and linear HMMs  structures have been considered and their properties investigated in the literature, see, e.g., \cite{Petreczky2007,Blackmore2007,Ozkan2013}, and references therein. 

In this paper, we identify the parameters of PWASS models. We assume that the boundaries of each region that form the piecewise function are known and use an approximate Bayesian smoother within the Expectation-Maximization (EM) algorithm to identify the model  parameters.
In our approach, the submodels are a function of the state, that is, given the state the active submodel is totally defined. The EM algorithm is used to identify the parameters of the submodels that form the PWASS model. In the E-step of the EM algorithm, we have used the cumulative distribution function (CDF) to compute the probability of each submodel given all available measurements.
 In the M-step, the parameters are estimated by maximizing a surrogate function for the likelihood function.

This paper is organized as follows; in section \ref{sec:probdef} is presented the model structure that we will be working with. In section \ref{sec:EM} the EM algorithm for PWASS is presented and discussed. In section  \ref{sec:ex} the EM algorithm is used to identify a PWASS model. Finally, the concluding remarks are given in section \ref{sec:concl}.

\section{Problem definition}
\label{sec:probdef}

Consider the following SSM
\begin{subequations}
\label{eq:nlss}
\begin{align}
\x_{t+1}=\ &\F(\x_t)+B \bu_t + \w_t,\\
\y_t=\ &C\x_t+\boldsymbol\nu_t,
\end{align}
\end{subequations}
where the state vector $\bfx_t\in \mathbb{R}^{n_{x}}$ is  partitioned by two scalar variables $\eta_t$ and $\zeta_t$, and a vector $\boldsymbol \chi_t\in \mathbb{R}^{(n_{x}-2)}$ such that $\x_t\triangleq \begin{bmatrix} \eta_t, \zeta_t, \boldsymbol \chi_t^\t \end{bmatrix}^\t$; $\bfy_t\in \mathbb{R}^{n_y}$ is the measurement; $C \in \mathbb{R}^{n_y\times n_x}$ is the known measurement matrix; $B\in \mathbb{R}^{n_x \times n_u}$ is the input matrix, $\bfu_t\in \mathbb{R}^{n_u}$ is the input. The initial state has a prior distribution $	\mathbf{\bfx}_1 \sim \mathcal{N}(\hat \bfx_{1|0},P_{1|0})$, where the subscript ``$t_1|t_2$" is read ``at time $t_1$ using measurements up to time $t_2$", and $\N(\boldsymbol \mu, \Sigma)$ means a Gaussian distribution with mean $\boldsymbol \mu$ and covariance $\Sigma$.  The process noise $\{\w_t\in\mathbb{R}^{n_x}| 1\leq t \leq T\}$  and the measurement noise $\{\boldsymbol \nu_t\in\mathbb{R}^{n_y}| 1\leq t \leq T\}$ are mutually independent Gaussian  noise sequences. 
The nonlinear function $\F(\cdot)$ is the state transition matrix given by
\begin{align}
\F(\x_t)\triangleq \begin{bmatrix}  \Phi^\t \x_t \\ \boldsymbol \phi^\t \z_t  +f(\eta_t)\\ F\x_t   \end{bmatrix}, \label{eq:pw_geral}
\end{align}
with $\Phi \in \mathbb{R}^{n_x}$, $F \in \mathbb{R}^{(n_x-2) \times n_x}$, $\z_t\triangleq \begin{bmatrix}  \zeta_t, \boldsymbol \chi_t^\t \end{bmatrix}^\t$, $\boldsymbol \phi \in \mathbb{R}^{(n_x-1)}$, and $f(\eta_t)$ is a piecewise affine function such that
\begin{equation}
f(\eta_t) = 
 \begin{cases}
   f_1 = a_1 {\eta}_{t} + b_1 & \text{if } {l}_{1}<{\eta}_{t}\leq{l}_{2}, \\
	\quad \vdots & \\
	  f_{N_r} =  a_{N_{r}} {\eta}_{t} + b_{N_{r}} & \text{if } {l}_{{N_{r}}}<{\eta}_{t}<{l}_{{N_{r}+1}}, \\
  \end{cases}
\label{eq:PWregion}
\end{equation}	
where ${l}_{i}, ~ i=1,\dots,N_r+1$ are the boundaries of each region that form the piecewise function. 
Rewriting \eqref{eq:pw_geral} for a region $\mathcal R_i \triangleq \left\{\eta_t :  {l}_{i}<{\eta}_{t}\leq{l}_{i+1}\right\}$ using \eqref{eq:PWregion}  we obtain 
\begin{align}
\F(\x_t) &=  \begin{bmatrix} 
 \Phi^\t \x_t   \\ \boldsymbol \phi^\t \z_t  +a_i {\eta}_{t} + b_i \\F\x_t 
\end{bmatrix} \\
&= \overbrace{ 
\begin{bmatrix} 
\Phi^\t  \\ [a_i ~\boldsymbol \phi^\t  ] \\ F  
\end{bmatrix}}^{A_i}\x_t  + 
\overbrace{
\begin{bmatrix} 
\mathbf{0} \\ b_i  \\ \mathbf{0}   
\end{bmatrix}}^{\mathbf{b}_i} \label{eq:pw_geral22} \\  
&=  A_i\x_t + \mathbf{b}_i .
\label{eq:pw_geral2}
\end{align}
Therefore, for a given  region $\mathcal R_i$, the model \eqref{eq:nlss} is a conditionally affine SSM
\begin{subequations}
\label{eq:nlss2}
\begin{align}
\x_{t+1}=\ &A_i \x_t+B \bu_t +\mathbf b_i + \w_t,\\
\y_t=\ &C\x_t+\boldsymbol\nu_t.
\end{align}
\end{subequations}
The index $i \in \left\{1,\cdots, N_{r}\right\}$ determines  which piecewise affine dynamic is activated at time $t$.
We will assume that the matrix $C$ is such that a direct measurement of the variable $\eta_t$ (up to a scalar product) is available in the measurement vector $\bfy_t$ and it is denoted by $y_t$.

Assuming that the piecewise function is continuous, that means,
\begin{equation}
a_i l_{i+1}+ b_i = a_{i+1} l_{i+1}+ b_{i+1} , \quad i=1,\dots, N_r-1,
\end{equation}
we can write each one of the $f_i$'s in \eqref{eq:PWregion} as
\begin{align}
f_1 &= a_1 \eta_t + b_1,\\
f_i &={a_i} \eta_t + b_i, \quad i=2,\dots, N_r,
\label{eq:general}
\end{align}
where
\begin{equation}
b_i = - a_i l_i  + {b_1 + a_1 l_1  + \sum_{j=1}^{i-1}{a_j (l_{j+1} - l_j)}}, 
\label{eq:bi}
\end{equation}

The identification problem consists of finding the parameters  
\begin{equation}
  \theta \bydef \left\{\left\{ a_i\right\}_{i=1}^{N_{r}}, \left\{ b_i\right\}_{i=1}^{N_{r}}, F, \Phi,\boldsymbol \phi\right\}
\label{eq:theta_vec}
\end{equation}
based on $\bfy_{1:T}$ (and $\bfu_{1:T}$), where $\bfy_{1:T}$ is a collection of $T$ observations. For the continuous piecewise function, using \eqref{eq:bi} the identification problem will be to find 
\begin{equation}
  \theta^{\prime} = \left\{\left\{ a_i\right\}_{i=1}^{N_{r}},b_1, F, \Phi, \boldsymbol \phi\right\}.
\end{equation}


\begin{figure}[tb]
\centering
 \includegraphics[width=1\columnwidth]{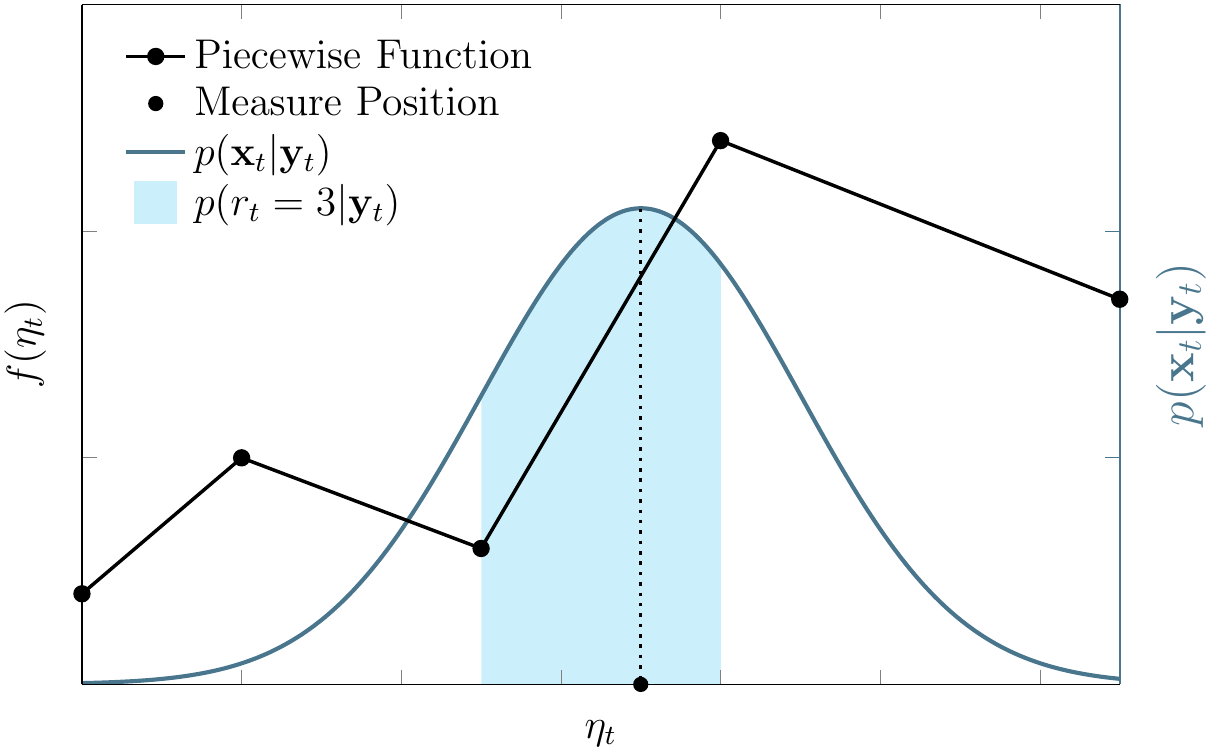}
\caption{The shaded area represents the probability that the system be in determinate region delimited by the piecewise function.}%
\label{fig:mode_prob}%
\end{figure}

\section{Solution using EM}
\label{sec:EM}
\begin{table}[t]
\caption{E-step of EM algorithm}\label{alg:estep}
\vspace{-1mm}\rule{\columnwidth}{1pt}
\begin{algorithmic}[1]
\State \textbf{Inputs:} $ A_{i},\mathbf b_{i}, l_i,  ~ i=1\dots, N_r$ , $B$, $C$, $Q$, $R$, $\mathbf u_{1:T}$, $\y_{1:T},\m_{1|0},P_{1|0}$
\For{$j = 1 $ to $M$}
\State $\m_{1|1} \gets \m_{1|0}$
\State $P_{1|1} \gets P_{1|0}$
\For{$t = 1 $ to $T-1$}
\Statex \textit{Sample Submodel Trajectory $r_{1:T}$}
\State $r_t^{(j)} \sim \int_{l_i}^{l_{i+1}}{\N(\eta_t;y_t,[R]_{(1,1)}) \d\eta_t}, \quad i=1,\dots, N_r$	
	\Statex \textit{Kalman filter prediction step}
		\State {
		$\widehat{\x}_{t+1|t} \gets A_{r_t^{(j)}} \m_{t|t}+B \bu_t +\mathbf b_{r_t^{(j)}} $
				}
		\State {  $P_{t+1|t}  \gets A_{r_t^{(j)}} P_{t|t} A_{r_t^{(j)}}^\t+Q  $}
		\Statex \textit{Kalman filter update step}
				\State $\Sigma_{t+1} \gets C P_{{t+1}|t} C^T + R$
		\State $K_{t+1} \gets  P_{{t+1}|t} C^T\Sigma^{-1}_{t+1}$
		\State {  $ 	\widehat{\x}_{{t+1}|{t+1}}   \gets \m_{{t+1}|t} + K_{t+1} (\mathbf{y}_{t+1} - C \m_{{t+1}|t}) $}					
		\State $P_{{t+1}|{t+1}} \gets P_{{t+1}|t} - P_{{t+1}|t}C^T \Sigma^{-1}_{t+1} C P_{{t+1}|t}$
				\EndFor

		\Statex \textit{RTS smoother}
		
		\State $P_{T|1:T} \gets P_{T|T}$
		\State $\m_{T|1:T} \gets \m_{T|T}$
		\For{$t = T-1$ to $1$}
		\State $ {G}_{t} \gets P_{t|t} A_{r_t}^T [{P}_{t+1|t}]^{-1}$
		\State $\hat{P}_{t,t-1|T}^{(j)} \gets  \hat{P}_{t+1|T}G_t^T $
		\State $\widehat{\bfx}_{t|1:T}^{(j)} \gets \widehat{\bfx}_{t|t} + {G}_{t}(\hat{\bfx}_{t+1|1:T}-\widehat{\bfx}_{t+1|t})$
		\State $ \hat{P}_{t|1:T}^{(j)} \gets {P}_{t|t} + {G}_{t}(\hat{P}_{t+1|1:T} - {P}_{t+1|t} ){G}_{t}^T$
\EndFor
\State $r_{1:T}^{(j)} \gets \left\{r_t^{(j)}\right\}, t=1,\dots,T$
	\EndFor
	\State \textbf{Outputs: $r_{1:T}^{(j)}, \hat{P}_{t|1:T}^{(j)}, \widehat{\bfx}_{t|1:T}^{(j)}, \hat{P}_{t,t-1|T}^{(j)},  j=1,\dots,M$ } 
\end{algorithmic}
\noindent \rule{\columnwidth}{1pt}\vspace{0mm}
\end{table}

The EM algorithm is an iterative method which is useful for approaching maximum likelihood estimate  of unknown parameter $\theta$ in probabilistic models involving latent variables defined by
	\begin{equation}
	\hat{\theta}_{ML} =  \operatorname*{arg\,max}_{\theta\in \Theta} ~ {p_\theta(\bfy_{1:T}|\theta)}.
	\label{eq:theta_ml}
	\end{equation}
The main idea of the EM algorithm \cite{Dempster1977a} is to compute an auxiliary function $\mathbf{Q}(\theta,\hat{\theta}_k)$ as a surrogate for the likelihood ${p_\theta(\bfy_{1:T}|\theta)}$.
In the expectation step (E-step) of the EM  approach,  first the following expectation should be computed
\begin{equation}
\begin{aligned}
&\mathbf{Q}(\theta,\hat{\theta}_k) =\operatorname{E}_{\hat{\theta}_k}\left[\log{p( \bfx_{1:T}, \bfy_{1:T} | \theta)|\bfy_{1:T}}\right],
\end{aligned}
\label{eq:EM_likelihood_original}
\end{equation}
where $\operatorname{E}_{\hat{\theta}_k}[\cdot]$ denotes the expectation with respect to  the latent variable whose posterior is computed using the previous estimate of $\theta$ denoted by $\hat{\theta}_k$.
However, the joint distribution $p( \bfx_{1:T}, \bfy_{1:T} | \theta)$ can not be computed analytically due to the nonlinearity of the state transition function. As a remedy, we use an approximation which uses a latent variable representing the submodel at each time step denoted by $r_t$ and with the probability density function (PDF)  
\begin{equation}
p(r_t)= \iint_{\eta_t\in \R_i}{p(\bfx_t) \d\eta_t \d\z_t}.
\end{equation}
Hence, $r_t \in \left\{1, 2, \cdots,{N_{r}}\right\}$ is a categorical random variable. We define $r_{1:t} \bydef \left\{r_1,\dots, r_t \right\}$ as the submodel trajectory up to time $t$. Also, we can compute the posterior distribution of $r_t$ given the measurement $\bfy_t$ via the following marginalization
\begin{align}
p(r_t|\bfy_t) \propto & \iint_{\eta_t\in \R_i}{p(\bfy_t|\bfx_t)p(\bfx_t) \d\eta_t \d\z_t}.
\end{align}
Assuming an uninformative prior on $\bfx_t$ and the fact that $\eta_t$ is directly measured, where its corresponding measurement is denoted by $y_t$, it gives the following categorical posterior distribution for $r_t$
\begin{align}
p(r_t|\bfy_t)  &=p(r_t=i|\bfy_t), \quad i=1,\cdots, N_r\nonumber\\
&= \int_{\eta_t\in \R_i}{\N(\eta_t;y_t,[R]_{(1,1)}) \d\eta_t}, \quad i=1,\cdots, N_r
\end{align}
where $[R]_{(1,1)}$ is the element of the first row and first column of $R$. 
  Fig. \ref{fig:mode_prob} shows the PDF $p(\x_t|\y_t)$ and a piecewise function with four regions. The shaded area represents the probability that the system be in  region three given the measurement $\y_t$, \ie, $p(r_t=3|\y_t)$. 
We will use the joint distribution  ${p_\theta(\x_{1:T}, \bfy_{1:T},r_{1:T}|\theta)}$ instead of  ${p_\theta(\x_{1:T}, \bfy_{1:T}|\theta)}$ in \eqref{eq:EM_likelihood_original}. This joint density is given by
\begin{align}
&p(\bfx_{1:T}, \bfy_{1:T}, r_{1:T}| \theta)  \nonumber\\
 &=p(\bfx_{1:T} | r_{1:T}, \theta) p(\bfy_{1:T} | \bfx_{1:T}, \theta) p(r_{1:T}| \bfy_{1:T},\bfx_{1:T}) \label{eq:jointlikelihoodexact}
\end{align}
However, we notice that $r_t|\bfx_t$ is deterministic \ie, given $\bfx_t$, $r_t$ is totally determined. As a remedy we use $p(r_{1:T}| \bfy_{1:T}) $ instead of $p(r_{1:T}| \bfy_{1:T},\bfx_{1:T})$ in \eqref{eq:jointlikelihoodexact}.
Since the transition density $p(\bfx_{t}|\bfx_{t-1},r_{t-1},\theta)$ is known to be ${\N(\bfx_{t}|A_{r_{t-1}}\bfx_{t-1} +B\bfu_{t-1} + \mathbf{b}_{r_{t-1}}, Q)}$, the joint distribution $p( \bfx_{1:T}, \bfy_{1:T} ,r_{1:T}| \theta)$ can be approximated  as in
\begin{align}
p( &\bfx_{1:T}, \bfy_{1:T}, r_{1:T}| \theta)  \nonumber\\
 &\approx p(\bfx_{1:T} | r_{1:T}, \theta) p(\bfy_{1:T} | \bfx_{1:T}, \theta) p(r_{1:T}| \bfy_{1:T}) \\
&= p(\bfx_1)p(\bfy_1|\bfx_1) p(r_1|\bfy_1) \nonumber\\
& \times \prod_{t=2}^{T} {p(\bfx_t|\bfx_{t-1},r_{t-1},\bfy_{t-1},\theta)p(r_{t}|\bfy_{t})p(\bfy_t|\bfx_t)},
\label{eq:likelihood_completa}
\end{align}
where
$
{p(\bfy_t|\bfx_{t})} = {\mathcal{N}(\bfy_t;C\bfx_t, R)}.
$
Hence, in the E-step of the EM  approach we calculate the expected value of the  log-likelihood $p(r_{1:T}, \bfx_{1:T}, \bfy_{1:T} | \theta)$ with respect to the observed data and the previous estimate $\hat{\theta}_k$: 
\begin{align}
&\mathbf{Q}(\theta,\hat{\theta}_k) = \operatorname{E}_{\hat{\theta}_k}\left[\log{p( \bfx_{1:T}, \bfy_{1:T},r_{1:T} | \theta)|\bfy_{1:T}}\right] \\
&= \iint{\log p( \bfx_{1:T}, \bfy_{1:T} ,r_{1:T}| \theta) } \nonumber \\ 
& \qquad \times p(\bfx_{1:T},r_{1:T} | \bfy_{1:T}, \hat{\theta}_k) \d\bfx_{1:T}\d r_{1:T},
\label{eq:EM_expectation}
\end{align}
then in the M-step the parameters are estimated by
	\begin{equation}
	\hat{\theta}_{EM} =  \operatorname*{arg\,max}_{\theta\in \Theta} ~ \mathbf{Q}(\theta,\hat{\theta}_k).
	\label{eq:theta_EM}
	\end{equation}

We notice that the only part that depends on $\theta$ in the joint posterior distribution \eqref{eq:likelihood_completa} is the term $p(\bfx_t|\bfx_{t-1},r_{t-1},\bfy_{t-1},\theta)$.  Thus the auxiliary quantity of the EM algorithm can be written as (omitting terms independent of $\theta$)
\begin{align}
&\mathbf{Q}(\theta,\hat{\theta}_k) =\operatorname{E}_{\hat{\theta}_k}\left[\log{ p(\bfx_{1:T}, \bfy_{1:T}, r_{1:T}, | \theta)|\bfy_{1:T}}\right]\\
=& \iint \sum^{T}_{t=2}{\left(\log{\mathcal{N}(\bfx_{t}|A_{r_{t-1}}\bfx_{t-1} +B\bfu_{t-1} + \mathbf{b}_{r_{t-1}}, Q)}\right)} \nonumber \\
&      \qquad \times p(\bfx_{1:T} ,r_{1:T}| \bfy_{1:T}, \hat{\theta}_k)\d\bfx_{1:T}\d r_{1:T},
\label{eq:EM_likelihood}
\end{align}
where  the first term of \eqref{eq:EM_likelihood} can be written as (omitting constant terms denoted by $\byplus$)  
\begin{multline}
\log{\mathcal{N}(\bfx_{t}|A_{r_{t-1}}\bfx_{t-1} +B\bfu_{t-1} + \mathbf{b}_{r_{t-1}}  , Q)} 
\\ \byplus \left\langle \Psi_{{r_{t-1}}}(\theta),s(\bfx_{t},\bfx_{t-1})\right\rangle + \Xi_{{r_{t-1}}}(\theta),
\end{multline} 
where $\left\langle a , b \right\rangle \bydef \text{tr}(a^T b)= a^T \cdot b$ denotes inner product and
\begin{align}
 \Psi_{{r_t}}(\theta) \bydef &
\begin{bmatrix}
 A_{r_t}^T Q^{-1}  \\ 
 \mathbf{b}_{r_t}^T Q^{-1}\\
 - \frac{1}{2} A_{r_t}^T Q^{-1}  A_{r_t}\\
 -     \left(B\mathbf{u}_{t-1} + \mathbf{b}_{r_t}\right)^T Q^{-1}A_{r_t}\\
\end{bmatrix}, \\
s(\bfx_{t},\bfx_{t-1})\bydef&
\begin{bmatrix}
\mathbf{x}_{t-1}  \mathbf{x}_{t}^T  \\
   \mathbf{x}_{t}^T \\
  \mathbf{x}_{t-1}\mathbf{x}_{t-1}^T  \\
  \mathbf{x}_{t-1}^T 
\end{bmatrix}.
\end{align}
Also, $\Psi_{{r_t}}(\theta)$ and  $s(\bfx_{t},\bfx_{t-1})$ are the natural parameter and the sufficient statistic, respectively. Further,
\begin{equation}
\Xi_{{r_t}}(\theta) \bydef - \mathbf{u}_{t-1}^T B^T Q^{-1} \mathbf{b}_{r_t}   - \frac{1}{2} \mathbf{b}_{r_t}^{T} Q^{-1} \mathbf{b}_{r_t}  
\end{equation}
 denotes the log-partition function. 
The auxiliary quantity of the EM algorithm can thus be written as (omitting constant terms) 
\small
\begin{align}
\mathbf{Q}&(\theta,\hat{\theta}_k) =\operatorname{E}_{\hat{\theta}_k}\left[\log{p( \bfx_{1:T}, \bfy_{1:T},r_{1:T} | \theta)|\bfy_{1:T}}\right]  \\ 
&\byplus \sum^{T}_{t=2} \left( \left\langle \Psi_{r_t}(\theta),\operatorname{E}_{\hat{\theta}_k}\left[s(\bfx_{t},\bfx_{t-1})| \bfy_{1:T}\right]\right\rangle + \operatorname{E}_{\hat{\theta}_k}\left[\Xi_{r_t}(\theta)| \bfy_{1:T}\right]  \right)  \\
&\byplus \sum^{T}_{t=2} \left\langle \Psi_{r_t}(\theta),\operatorname{E}_{\hat{\theta}_k}\left[s(\bfx_{t},\bfx_{t-1})| \bfy_{1:T}\right]\right\rangle + \sum^{T}_{t=2}\Xi_{r_t}(\theta).
\label{eq:Qem}
\end{align}
\normalsize
In order to complete the calculation of $\mathbf{Q}(\theta,\hat{\theta}_k)$, we need to evaluate the expected value in \eqref{eq:Qem} with respect to the observed data.  This amounts to calculating the integral in \eqref{eq:EM_likelihood}. 
As in \eqref{eq:likelihood_completa}, the second term of \eqref{eq:EM_likelihood} can be factorized as 
\begin{equation}
p(r_{1:T}, \bfx_{1:T} | \bfy_{1:T}, \hat{\theta}_k) = {p(\bfx_{1:T}|\bfy_{1:T},r_{1:T},\hat{\theta}_k)} {\prod_{t=1}^{T}{p(r_{t}| \bfy_{t}) }},
\label{eq:GaussSumSmother}
\end{equation}
where it was assumed that 
\begin{equation}
p(r_{1:T}|\y_{1:T})\approx \prod_{t=1}^T p(r_t|\y_t).
\end{equation}

 Since  $r_1, \cdots, r_t \in \left\{1, 2, \cdots,{N_{r}}\right\}$, the number of possible trajectories up to time $t$ is $N_r^t$, \ie, it grows exponentially with time. The integral \eqref{eq:EM_likelihood} can be computed using various integration methods. We use Monte-Carlo integration, where the samples are drawn from 
\begin{align}
r_t^{(j)} &\sim p(r_t | \bfy_t), \quad j=1,\cdots, M.
\label{eq:sample}
\end{align}
When the samples are drawn, the remaining integral can be computed analytically with the approximation of the posterior of each trajectory to be Gaussian. 
With such as assumption, we can use the Kalman Smoother also called  Rauch-Tung-Striebel (RTS) smoother for linear and Gaussian SSM \cite{Rauch1965}. The algorithm in Table \ref{alg:estep} presents the E-step of the proposed EM algorithm. The backward recursion equations for the RTS smoother are given in the lines 14-20 of  Table \ref{alg:estep}. 
 Using the RTS smoother we can write the expectation  $\operatorname{E}_{\hat{\theta}_k}\left[s(\bfx_{t},\bfx_{t-1})| r^{(j)}_{1:T}, \bfy_{1:T}\right]$ as
\begin{multline}
\operatorname{E}_{\hat{\theta}_k}\left[s(\bfx_{t},\bfx_{t-1})| r^{(j)}_{1:T}, \bfy_{1:T}\right]= \\
\overbrace{
\begin{bmatrix}
\hat{\bfx}_{t-1|1:T}^{(j)}(\hat{\bfx}_{t|1:T}^{(j)})^T + \hat{P}_{t,t-1|1:T}^{(j)}\\
\hat{\bfx}_{t|1:T}^{(j)}\\
\hat{\bfx}_{t-1|1:T}^{(j)}(\hat{\bfx}_{t-1|1:T}^{(j)})^T + \hat{P}_{t-1|1:T}^{(j)}\\
\hat{\bfx}_{t-1|1:T}^{(j)}\\
\end{bmatrix}}^{S^{(j)}_t},
\label{eq:suff_est_rt}
\end{multline}
 Here the superscript $^{(j)}$ means that the quantity is conditioned to the $j\mbox{-}\text{th}$ sampled trajectory $r_{1:T}$. Finally, inserting \eqref{eq:suff_est_rt} in \eqref{eq:Qem} we find
\begin{multline}
\mathbf{Q}(\theta,\hat{\theta}_k)  
=\frac{1}{M}\sum^{T}_{t=2}  \sum^{M}_{j=1}\left(  \left\langle \Psi_{r_t^{(j)}}(\theta),{S^{(j)}_t}\right\rangle + \Xi_{r_t^{(j)}}(\theta)\right).
\label{eq:Qem_rt2}
\end{multline}

In the M-step, it is possible to calculate exactly the gradient and the Hessian of \eqref{eq:Qem_rt2} with respect to $\theta$ and use them within the Newton method to find the maximum of $\mathbf{Q}(\theta,\hat{\theta}_k)$. The quantities used in \eqref{eq:Qem_rt2} are the outputs of the algorithm in Table \ref{alg:estep}.
It is worth pointing out that the computational complexity to calculate \eqref{eq:Qem_rt2} is $\mathcal{O}(TM).$

\section{Numerical Simulation}
\label{sec:ex}


\begin{figure*}[!th]
  \centering
 \includegraphics[width=1\textwidth]{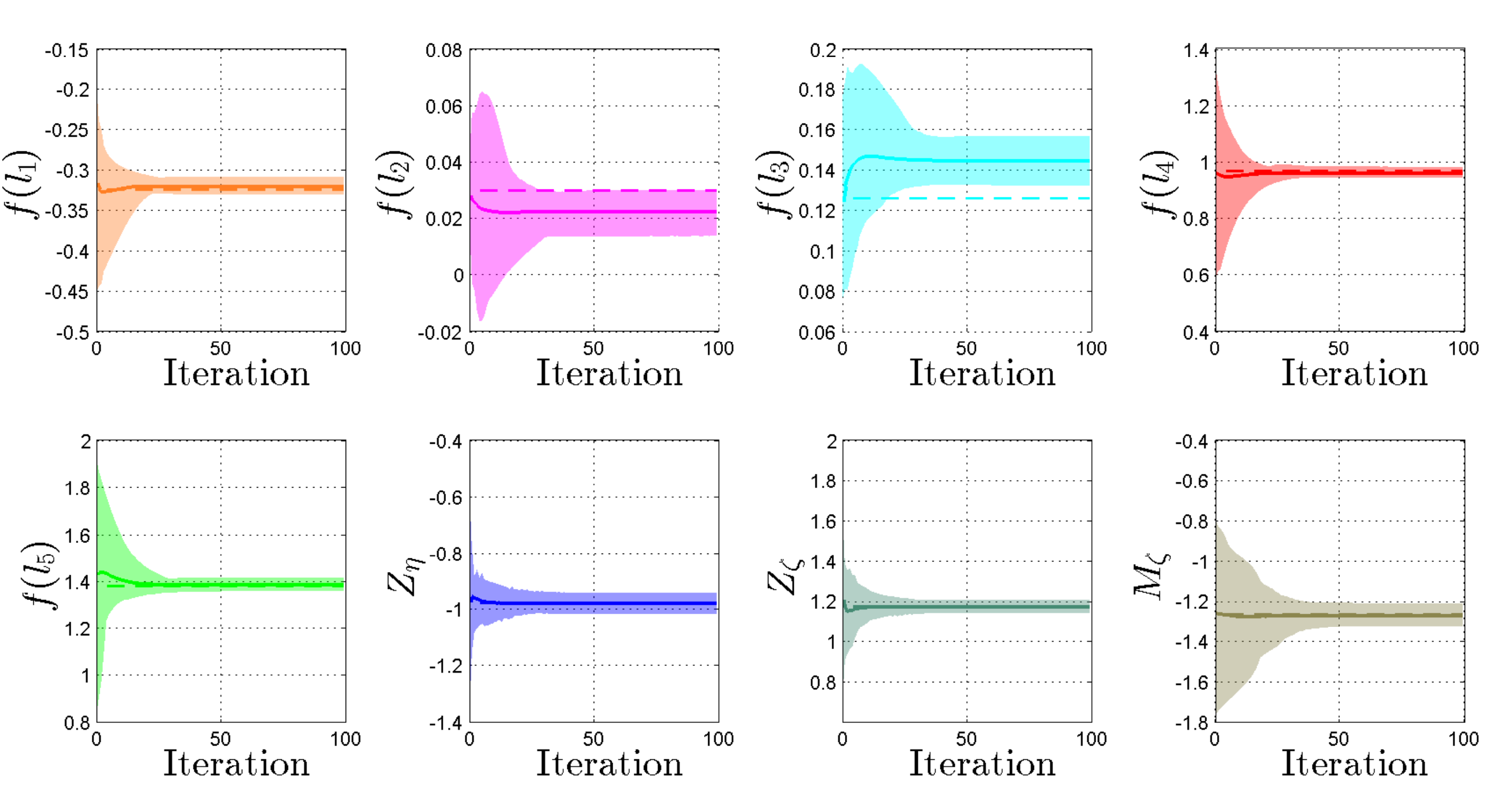}
  \caption{Parameter identification over 100 EM iterations. The lines show the averages (full line) and the true values of the parameters (dashed line), the transparent shaded areas show the upper and lower bounds over 150 independent runs for different parameter initialization.}
\label{fig:ex}
\end{figure*}

We will evaluate the methodology presented above through an example of the  identification of JAS 39 Gripen's flight dynamic in the longitudinal direction. The discretized SSM of the system is given by \cite{Larsson2013}
\begin{subequations}
	\begin{align}
	\begin{bmatrix}
		{\eta}_{t+1} \\ {\zeta}_{t+1}
	\end{bmatrix} &= 
		\begin{bmatrix}
		\eta_t + Z_\eta \eta_t + \zeta_t +  Z_\zeta \zeta_t \\ f(\eta_t) + M_\zeta \zeta_t  
	\end{bmatrix} \nonumber \\ 
		 & \qquad  \quad + \begin{bmatrix}
		Z_{\delta_e} &  Z_{\delta_c} \\ M_{\delta_e} & M_{\delta_c}
	\end{bmatrix} 		
	\begin{bmatrix}
		{\delta_e}_t\\ {\delta_c}_t
	\end{bmatrix} + \mathbf{w}_t,\\
	\mathbf{y}_t &= 		
	\begin{bmatrix}
		1 & 0  \\ 0 & 1 
	\end{bmatrix}
		\begin{bmatrix}
		{\eta}_t \\ {\zeta}_t
	\end{bmatrix}
	+ \boldsymbol\nu_t,
	\end{align}%
\label{eq:modelo}%
\end{subequations}%
where the process noise $\w_t$ and the measurement noise $\boldsymbol{\nu}_t$ are Gaussian with known mean and covariance, $\eta_t$ is the angle of attack, $\zeta_t$ is the pitch rate of the aircraft, ${\delta_e}_t$ and ${\delta_c}_t $ are the elevator and canard control action. 
The nonlinear function $f(\eta_t)$ is constructed as a continuous piecewise function with $N_{r}$ known regions as in \eqref{eq:general}.
The goal is to identify the parameters $Z_\eta = -0.9759 $, $Z_\zeta =   1.174 $, $M_\zeta = -1.2616$ and the piecewise function values
 $$ \left[\left\{ f(l_i)\right\}_{i=1}^{N_r+1}\right]= \left[-0.3240~ 0.0300~ 0.1260~ 0.9660~ 1.3800\right],$$ where the boundaries for each region $l = [l_1, \dots, l_5]$ are
$$l =[-1^\circ, 4^\circ, 7^\circ, 12^\circ, 16^\circ ].$$
The system is simulated for $T=1\:800$ time steps and the sample time is $\Delta t = 1/60s$. The input signal is such that all submodels are activated about the same amount of times. The parameters of the input matrix are: $Z_{\delta_e}= 0.3043 $, $  Z_{\delta_c} = 0.0289$ ,$ M_{\delta_e}= -31.0898$ and  $M_{\delta_c}= 8.2557$.
 It is worth mentioning that the system described in \eqref{eq:modelo} is unstable, so an LQ regulator is used \cite{Larsson2013}. This LQ regulator adds a correlation between the control signal and the measurement noise as well as state noise. However, this is ignored here.

\begin{figure}
\centering
   \includegraphics[width=0.85\columnwidth]{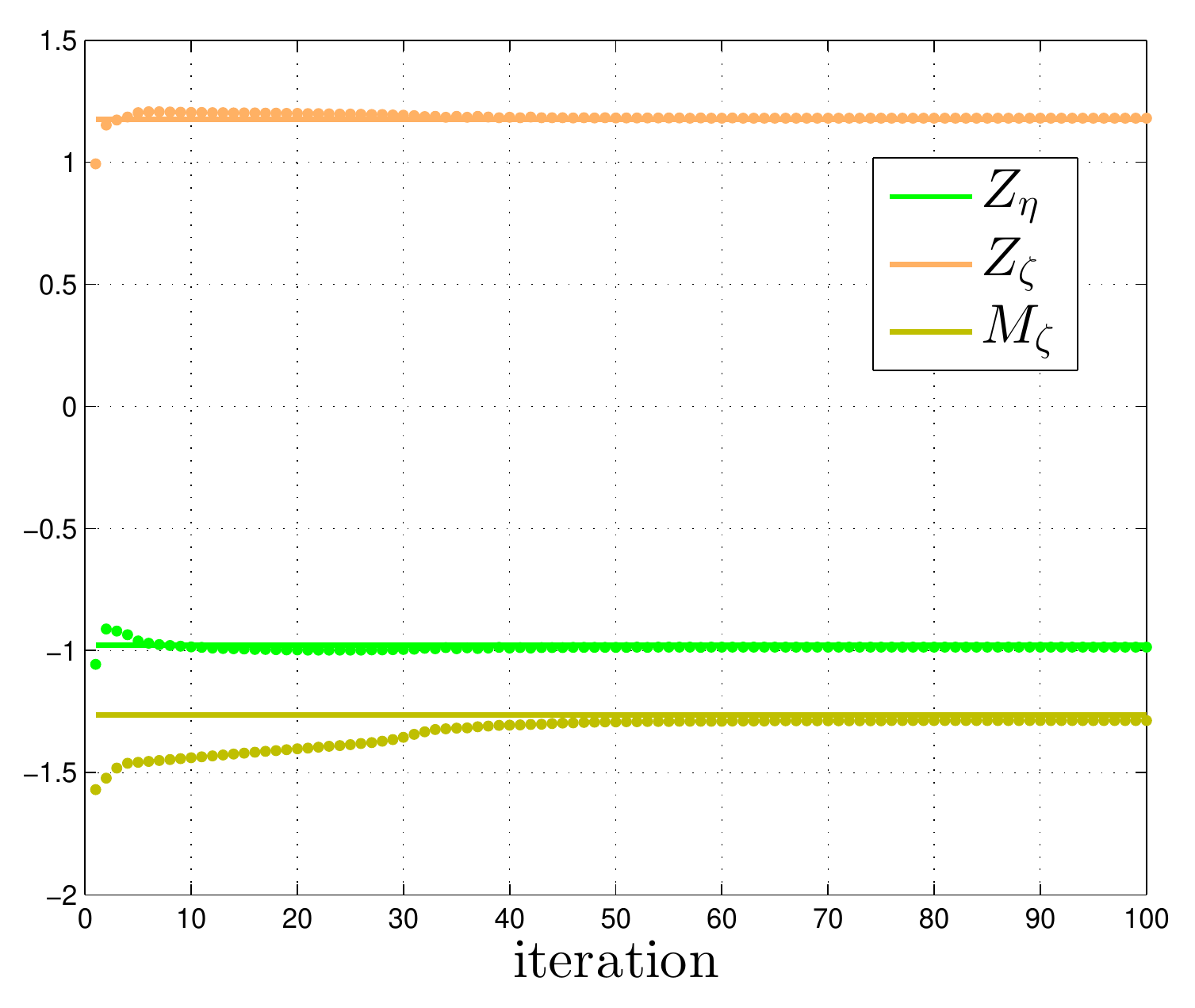}%
	\caption{True and estimate parameters $Z_\eta, Z_\zeta$ and $M_\zeta$ versus the EM iterations.}
   \label{fig:ex2_theta}
\end{figure}

\begin{figure*}
\centering
   \begin{subfigure}[b]{\columnwidth}
    \includegraphics[width=0.9\linewidth]{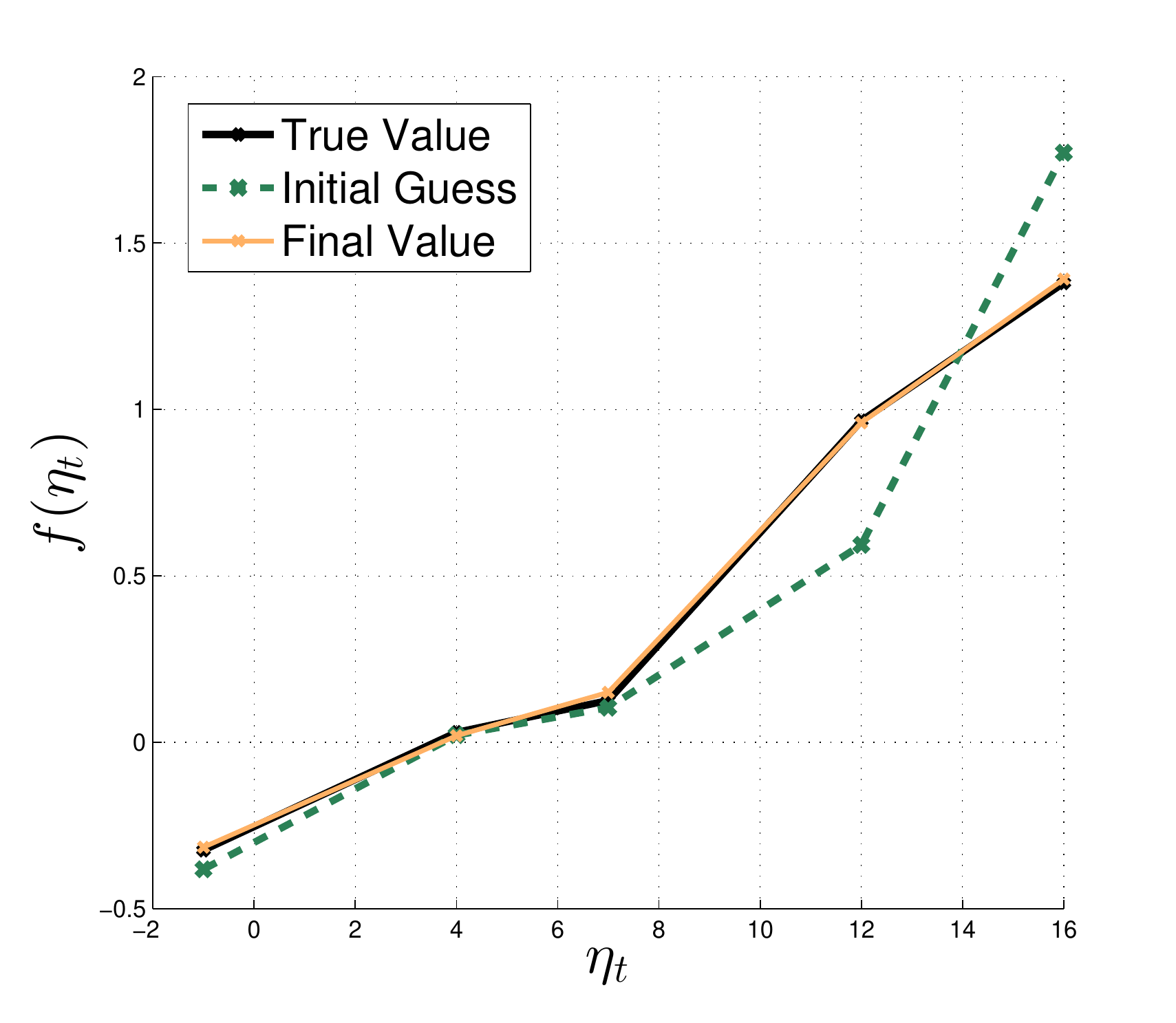}%
	\caption{}
   \label{fig:ex2_mode} 
\end{subfigure}
\begin{subfigure}[b]{\columnwidth}
    \includegraphics[width=0.9\linewidth]{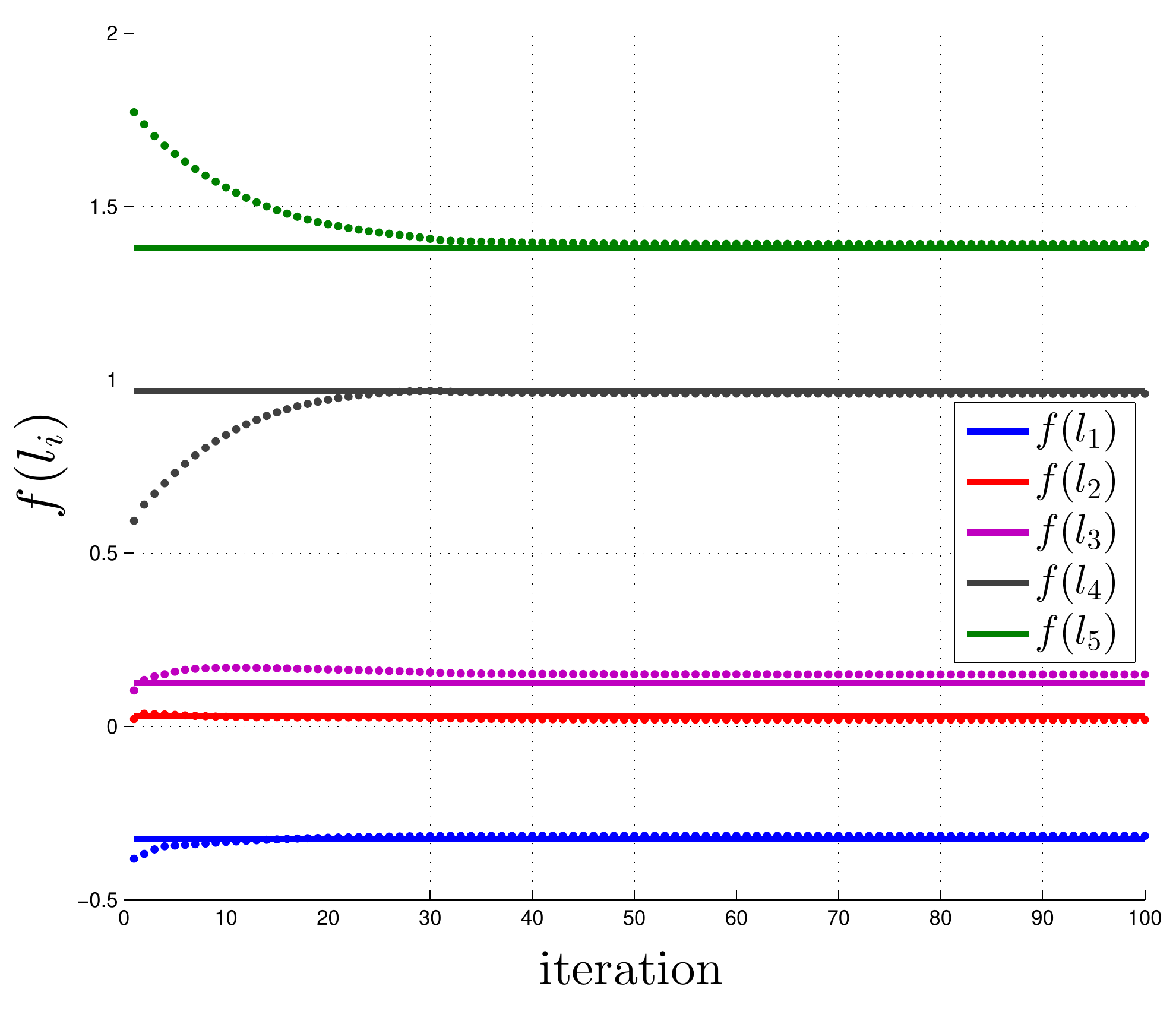}%
	\caption{}
   \label{fig:ex2_pw}
\end{subfigure}
\caption{Fig. \ref{fig:ex2_mode} : True, initial guess and final estimate of the piecewise function.  Fig. \ref{fig:ex2_pw}: True and estimated values of $f(l_i)$  versus the EM iterations.}%
\label{fig:ex2}
\end{figure*}

 We have used \eqref{eq:sample} to sample $M=300$ possible trajectories $r_{1:T}$ and evaluate the integral on the right-hand side of \eqref{eq:EM_likelihood}.  The MATLAB function $\operatorname{fminunc}$ is used to find the solution of \eqref{eq:theta_EM}.
 Fig. \ref{fig:ex} presents the identification of the model \eqref{eq:modelo} for 150 different realizations and 100 iterations of the EM algorithm. The shaded area in each figure represents the upper and lower bounds over the 150 different realizations. The initialization of $\theta$ is chosen randomly and uniformly, but such that each entry laid in an interval  equal to $40\%$ of the corresponding entry in the true parameter vector. We have set $\w_t \sim \N(0,\text{diag}[0.06^\circ,0.06^\circ])$, $\boldsymbol{\nu}_t \sim \N(0,\text{diag}[0.6^\circ,0.6^\circ])$  and $\x_1 \sim \N(0,\text{diag}[0.06^\circ])$. Fig. \ref{fig:ex2_theta} and \ref{fig:ex2} present one of those 150 realizations presented in the Fig. \ref{fig:ex}. The Fig. \ref{fig:ex2_mode} shows the true piecewise function, the initial guess, and the last estimate. In Fig. \ref{fig:ex2_pw} is presented the estimated value of the piecewise function calculated at each boundary of the function, \ie, $f(l_i),$  versus the EM algorithm iteration. In Fig. \ref{fig:ex2_theta}, the estimated values of the parameters $Z_\eta, Z_\zeta$, and $M_\zeta$ versus  the EM algorithm iteration are given.

 We notice that there is a bias in the final value of the parameters $f(l_2)$ and $f(l_3)$.
However,  it is possible to verify in Fig. \ref{fig:ex2_mode} that the estimated piecewise function can describe the true piecewise function.   In that case the estimated values were $\hat f(l_2)= 0.02001$ and $\hat f(l_3)= 0.1503$, equivalent to an error between the true value and the estimated value of, respectively, $0.33$ and $0.19$.

%


\section{Conclusion}
 \label{sec:concl}

We have proposed a method based on the EM algorithm for identification of PWASS models. We use the direct but noisy measure of the nonlinear state to calculate the probability for a given region for each time. The proposed EM algorithm was applied to the identification of the JAS 39 Gripen's flight dynamic in the longitudinal direction. In this example, a piecewise affine function with four regions was successfully identified as well as the remaining parameters of the state matrix. The results have shown that the proposed method can be used to identify PWASS models.   

\section{Acknowledgment}
The authors would like to thank Martin Enqvist and Roger Larsson for their inspirational role in this work. 

\small
\bibliographystyle{IEEETran}
\bibliography{IEEEabrv,library}

\end{document}